\documentclass{article}                     % onecolumn (standard format)
\usepackage{graphicx}
\usepackage{amssymb}
\usepackage{amsfonts}
\usepackage{amsmath}
\usepackage{multirow}
\usepackage{listings}
\usepackage[section]{placeins}

  \newcounter{example}[section]

\begin{document}

\title{Joining relations under discrete uncertainty}

%\subtitle{Do you have a subtitle?\\ If so, write it here}

%\titlerunning{Short form of title}        % if too long for running head

\author{Matteo Magnani, Danilo Montesi
\thanks{M.~Magnani is with the Department of Computer Science, Aarhus University, Denmark (magnanim@cs.au.dk), D.~Montesi is with the Department of Computer Science, University of Bologna, Italy (montesi@cs.unibo.it)}
}

%\authorrunning{Short form of author list} % if too long for running head

%\institute{Matteo Magnani \at
%              Department of Computer Science, University of Bologna\\
%              Tel.: +39-051-2094848\\
%              Fax: +39-051-2094983\\
%              \email{matteo.magnani@cs.unibo.it}           %  \\
%%             \emph{Present address:} of F. Author  %  if needed
%           \and
%           Danilo Montesi \at
%              Department of Computer Science, University of Bologna\\
%              Tel.: +39-051-2094887\\
%              Fax: +39-051-2094510\\
%              \email{montesi@cs.unibo.it}           %  \\
%}

%\date{Received: date / Accepted: date}
% The correct dates will be entered by the editor

\maketitle

\begin{abstract}
In this paper we introduce and experimentally compare alternative algorithms to join uncertain relations. Different algorithms are based on specific principles, e.g., sorting, indexing, or building intermediate relational tables to apply traditional approaches. As a consequence their performance is affected by different features of the input data, and each algorithm is shown to be more efficient than the others in specific cases. In this way statistics explicitly representing the amount and kind of uncertainty in the input uncertain relations can be used to choose the most efficient algorithm.
% \PACS{PACS code1 \and PACS code2 \and more}
% \subclass{MSC code1 \and MSC code2 \and more}
\end{abstract}

%\IEEEkeywords{Uncertainty, Join algorithms, Relational model}

%%%%%%%%%%%%%%%%%%%%%%%%%%%%%%%%%%%%%%%%%%%%%%%%%%%%%%%%%%%%%%%%%%%%%%%%%%%%%%%%%%%%%%%%%%%%%%%%%%%%%%%%%%%%%%%%%%%%%%
%%%%%%%%%%%%%%%%%%%%%%%%%%%%%%%%%%%%%%%%%%%%%%%%%%%%%%%%%%%%%%%%%%%%%%%%%%%%%%%%%%%%%%%%%%%%%%%%%%%%%%%%%%%%%%%%%%%%%%
%%%%%%%%%%%%%%%%%%%%%%%%%%%%%%%%%%%%%%%%%%%%%%%%%%%%%%%%%%%%%%%%%%%%%%%%%%%%%%%%%%%%%%%%%%%%%%%%%%%%%%%%%%%%%%%%%%%%%%
\section{Introduction}

One fundamental step of a traditional relational query optimization process is the choice of the best algorithm to execute each query operator. This is particularly important with regard to join operators, for which naive executions are computationally impractical and efficient algorithms have been developed. However, traditional join algorithms cannot be directly applied when the underlying model is extended to accommodate uncertain data.

In this paper we extend traditional join algorithms to work with uncertain data and discuss the outcomes of their experimental comparison. As a result we show that no algorithm is always more efficient than the others and we study their behavior with respect to parameters (statistics) that can be used to choose the best one depending on the input data.

In particular, we will focus on what is often called \emph{discrete} uncertainty. As we will detail in Section~\ref{related} the design of a join algorithm depends on the underlying model used to represent uncertainty. Discrete uncertainty models assume to have finite sets of alternative values instead of single certain values, e.g., the set \{45,46,47\} to represent the (partially unknown) age of an individual, and are currently implemented in the main uncertain data management systems like Trio \cite{Widom05} or Orion \cite{Cheng05}, where it is also possible to specify continuous uncertainty distributions.

In the following we introduce the problem of joining uncertain relations by example showing why traditional methods cannot be applied, and list the main contributions of this work. Then we present an overview of existing works on the topic. In Section~\ref{algo} we introduce the algorithms analyzed in the paper, in Section~\ref{exp} we report on their experimental comparison and in Section~\ref{interpretation} we provide an interpretation of the results of the experimental evaluation. We conclude the article with a summary and discussion of the main results.

\subsection{A brief recall of traditional join approaches and why they do not work on uncertain data}

Consider the relational table illustrated in Figure~\ref{fig:forbes}, representing the wealth of some notable American people. In this paper we focus on the equi-join operator, i.e., a join where tuples are concatenated and included in the result when some of their attributes have the same value. For example, we may want to join this table with itself on the \texttt{Net Worth} attribute to pair people with the same wealth:
\begin{tabbing}
SELECT p1.surname, p1.networth,\\
%\>\>
p2.surname, p2.networth\\
%\>
FROM WEALTH p1, WEALTH p2\\
%\>
WHERE p1.networth = p2.networth
\end{tabbing}
The result of this query is illustrated in Figure~\ref{fig:exresult}.

% www.articlesbase.com/wealth.../the-wealthiest-people-in-america-1715213.html
% Forbes - wikipedia
% Rich bastards
% http://www.richpeople.org/
\begin{figure}
\begin{center}
\begin{tabular}{|c||l|l|l|l|l|}
\hline
xid & \textbf{Name} & \textbf{Surname} & \textbf{Net Worth} & \textbf{Age} \\
\hline \hline
t1 & William H. & Gates III & 53 & 54 \\
\hline
t2 & Warren & Buffett & 47 & 79 \\
\hline
t3 & Paul & Allen & 14 & 57 \\
\hline
t4 & Lawrence & Page & 18 & 37 \\
\hline
t5 & Lawrence & Ellison & 28 & 65 \\
\hline
t6 & Michael & Dell & 14 & 45 \\
\hline
\end{tabular}
\end{center}
\caption{A table with the wealth of some notable American people in Billion USD (source: Forbes)}
\label{fig:forbes}
\end{figure}

% www.articlesbase.com/wealth.../the-wealthiest-people-in-america-1715213.html
% Forbes - wikipedia
% Rich bastards
% http://www.richpeople.org/
\begin{figure}
\begin{center}
\begin{tabular}{|l|l|l|l|}
\hline
\textbf{Surname} & \textbf{Net Worth} & \textbf{Surname} & \textbf{Net Worth} \\
\hline \hline
 Gates III & 53 &  Gates III & 53\\
\hline
 Buffett & 47 &  Buffett & 47\\
\hline
 Ellison & 28 &   Ellison & 28\\
\hline
 Page & 18 &  Page & 18\\
\hline
Allen & 14 &  Allen & 14\\
\hline
Allen & 14 &  Dell & 14 \\
\hline
 Dell & 14 &  Dell & 14 \\
\hline
 Dell & 14 &  Allen & 14\\
\hline
\end{tabular}
\end{center}
\caption{Result of a self-join}
\label{fig:exresult}
\end{figure}

A well known naive approach to obtain this result, the one available on the early relational systems and known as \emph{nested loop join}, consists in comparing each row of the first table (called \emph{external} relation) with all rows of the other (\emph{internal}): in our working example we would start comparing row t1 against t1, then (t1, t2), (t1, t3) and so on.

In Figure~\ref{fig:nestedloop} we have represented the order of comparisons between tuples and the found matches. Evidently this method
requires $n_1 \times n_2$ tuple comparisons where $n_1$ and $n_2$ are the cardinalities of the two input relations (36 comparisons in our example). Although potentially useful for small datasets or as a subroutine of more sophisticated join methods, this behavior does not scale to large datasets.

\begin{figure}
\begin{center}
\begin{tabular}{r||c|c|c|c|c|c|}
  & t1 & t2 & t3 & t4 & t5 & t6 \\
\hline \hline
t1 & \textbf{1}  & 2 & 3  & 4 & 5 & 6 \\
\hline
t2 & 7  & \textbf{8}  & 9  & 10  & 11  & 12  \\ 
\hline
t3 & 13  & 14 & \textbf{15} & 16 & 17  & \textbf{18}  \\ 
\hline
t4 & 19  & 20 & 21  & \textbf{22}  &  23 & 24 \\ 
\hline
t5 & 25  & 26  & 27 & 28  & \textbf{29} & 30\\ 
\hline
t6 & 31  & 32  & \textbf{33}  & 34  & 35 & \textbf{36}\\ 
\hline
\end{tabular} \hspace{.1cm}
\end{center}
\caption{Comparisons performed using a nested loop join. Numbers indicate the order of comparisons and boldface the matches found}
\label{fig:nestedloop}
\end{figure}

To reduce the number of comparisons several approaches have been proposed and are currently used in relational systems. All approaches are based on re-organizing the data such that for every row of one table there is no need to scan all the other table but we can look at specific locations where potentially matching records have been collected --- this is usually done by sorting the tables, using hash functions, or building indexes on the join columns. In this way it is not necessary to check all the other rows. For example, consider Figure~\ref{fig:sortjoinnounc}: here the input data have been sorted on the join attribute. When we compare row t6 (Michael Dell \textbf{14} 45) against row t4 (Lawrence Page \textbf{18} 37), which is the third comparison performed by this algorithm as indicated in the figure, we know that we will find no other matches, because all the following tuples in the internal input relation will have a value equal or greater than 18 and thus cannot match 14. As a consequence, we can proceed to row t5 without checking all other tuples in the second relation.

Finally, Figure~\ref{fig:index} shows the comparisons performed using an index. Here we can use the index to obtain directly the pointers to the matching tuples, if present, but we need to access the index for each row in the external relation --- we will further discuss the impact of this approach when applied to uncertain data in the section on the experimental comparison of the algorithms.

\begin{figure}
\begin{center}
\begin{tabular}{r||c|c|c|c|c|c|}
  & t6 & t5 & t4 & t3 & t2 & t1 \\
\hline \hline
t6 & \textbf{1}  & \textbf{2} & 3  &  &  &  \\
\hline
t3 & \textbf{4}  & \textbf{5}  & 6  &   &   &   \\ 
\hline
t4 & 7  &  8 & \textbf{9}  &  10 &   &   \\ 
\hline
t5 &   &   & 11  & \textbf{12}  & 13  &   \\ 
\hline
t2 &   &   &   & 14 & \textbf{15}  & 16  \\ 
\hline
t1 &   &   &   &   & 17  &  \textbf{18} \\ 
\hline
\end{tabular} \hspace{.1cm}
\end{center}
\caption{Comparisons performed using a sort join algorithm. Notice that the input tuples have been sorted on the join attribute}
\label{fig:sortjoinnounc}
\end{figure}

\begin{figure}
\begin{center}
\begin{tabular}{r||c|c|c|c|c|c|}
 & t1 & t2 & t3 & t4 & t5 & t6 \\
%\cline{1-1} \cline{3-8}
\hline \hline
INDEX(t1) $\rightarrow$ & \textbf{1} &  &  & & & \\
%\cline{1-1} \cline{3-8}
\hline
INDEX(t2) $\rightarrow$ & & \textbf{2}  &  & & & \\
%\cline{1-1} \cline{3-8}
\hline
INDEX(t3) $\rightarrow$ & & & \textbf{3} & & & \textbf{4}\\
%\cline{1-1} \cline{3-8}
\hline
INDEX(t4) $\rightarrow$ & & & & \textbf{5} & & \\
%\cline{1-1} \cline{3-8}
\hline
INDEX(t5) $\rightarrow$ & & & & & \textbf{6} & \\
%\cline{1-1} \cline{3-8}
\hline
INDEX(t6) $\rightarrow$ & & & \textbf{7} & & & \textbf{8}\\ 
%\cline{1-1} \cline{3-8}
\hline
\end{tabular} \hspace{.1cm}
\end{center}
\caption{Comparisons performed using an index join. Matching tuples are identified directly, but every tuple in the external relation requires an access to the index. In this example we are using only the index on one of the two input relations}
\label{fig:index}
\end{figure}

Now assume that the table has been filled using a Web information extraction tool, or even a manual browsing of different Web sites. Given the uncertainty of the data generation process (first case) or the inconsistency of on-line information (second case) the collected data would be uncertain. As a working example in Figure~\ref{fig:forbes-unc} we have represented the same table of Figure~\ref{fig:forbes} augmented with uncertain data coming from alternative Web sites.

Uncertainty is a state of limited knowledge about a past, current or future state of the world. Usually, uncertainty is represented using a set of alternative information items, e.g., \{18,19\} to represent the fact that the correct wealth of a person is either 18 or 19 million USD. In addition numerical or symbolic values are associated to these alternative elements, e.g., to indicate their probability or degree of possibility. However, in the following we use a possible world data model without numerical uncertainty measures. While computing a join, we first look for pairs of tuples to be concatenated and included in the result, and our algorithms aim at reducing the number of comparisons and I/Os. In this way, they can be used independently of the adopted uncertainty management theory --- probabilities, possibilities or other measures can be computed after the generation of the result.

\begin{figure}
\begin{center}
\begin{tabular}{|c||l|l|l|l|l|}
\hline
xid & \textbf{Name} & \textbf{Surname} & \textbf{Net Worth} & \textbf{Age} \\
\hline \hline
ut1 & William H. & Gates III & \{53,50,40,58\} & 54 \\
\hline
ut2 & Warren & Buffett & \{47,40,37,42\} & 79 \\
\hline
ut3 & Paul & Allen & \{14,16,22\} & 57 \\
\hline
ut4 & Lawrence & Page & \{18,19\} & 37 \\
\hline
ut5 & Lawrence & Ellison & \{28,23,25\} & 65 \\
\hline
ut6 & Michael & Dell & \{14,16,18\} & 45 \\
\hline
\end{tabular} \hspace{.1cm}
\end{center}
\caption{Our working example, containing a \texttt{NetWorth} uncertain attribute}
\label{fig:forbes-unc}
\end{figure}

Looking at the table in Figure~\ref{fig:forbes-unc} it appears why some efficient traditional methods can no longer be applied. For example, we can no longer sort its rows on the join attribute: both sets \{18,19\} (ut4) and \{14,16,22\} (ut3) contain values greater and smaller than the values of the other set. This motivates the extension of the aforementioned algorithms that will be presented in the next section and later experimentally evaluated.

\subsection{Main contributions}
In this article we provide the following main contributions:
\begin{itemize}
\item We define several algorithms to join uncertain relations. One algorithm is an extension of the relational sort join, and reduces to it when the data is certain. The other (tuple-based, presented in two variations) is an original algorithm that uses traditional joins between the simple tuples composing the uncertain relations. We also include in our evaluation an index-join, as suggested in other works, and two base algorithms (a nested loop approach and the one chosen by the underlying relational query optimizer used in our experiments) to be used as baselines. As we will see, except for the nested loop join no approach is always better than the others.
\item We compare the algorithms on several data sets to find the relationships between the input data and their performance.
\end{itemize}

The main results obtained after the experimental evaluation of the proposed algorithms show that:
\begin{itemize}
\item Tuple-based approaches, which transform uncertain tables into larger traditional relational tables with one tuple for each alternative value, have a time complexity which is independent of the distribution of uncertain values in the data, assuming a linear complexity of the underlying relational joins. However, their performance is affected by the number of alternative values contained inside each uncertain tuple.
\item When alternative values contained inside each uncertain tuple are similar, as it may be the case when we measure temperatures, lengths, etc., the sort-based approach is almost as efficient as a traditional sort-join. Therefore, this algorithm can be either more or less efficient than the tuple-based approach, depending on: the percentage of uncertain tuples, the number of alternative values for each uncertain field, the distribution (spreading) of these values, the cardinality of the input relations and the number of required tuples in the result in case it is not necessary to build all the resulting relation. 
\item The sort-based approach improves significantly when we require only the first few tuples of the result, as it usually happens in modern database system GUIs where additional tuples are fetched into memory only if explicitly required by the user. However, in this case using an index results to be the most efficient approach.
\item Using the most appropriate algorithm we are able to efficiently join tables containing millions of uncertain tuples.
\end{itemize}

%%%%%%%%%%%%%%%%%%%%%%%%%%%%%%%%%%%%%%%%%%%%%%%%%%%%%%%%%%%%%%%%%%%%%%%%%%%%%%%%%%%%%%%%%%%%%%%%%%%%%%%%%%%%%%%%%%%%%%
%%%%%%%%%%%%%%%%%%%%%%%%%%%%%%%%%%%%%%%%%%%%%%%%%%%%%%%%%%%%%%%%%%%%%%%%%%%%%%%%%%%%%%%%%%%%%%%%%%%%%%%%%%%%%%%%%%%%%%
%%%%%%%%%%%%%%%%%%%%%%%%%%%%%%%%%%%%%%%%%%%%%%%%%%%%%%%%%%%%%%%%%%%%%%%%%%%%%%%%%%%%%%%%%%%%%%%%%%%%%%%%%%%%%%%%%%%%%%
\section{Related work}\label{related}

Uncertain relational models have been studied since the early 90's, and the first works on this topic mainly focused on theoretical aspects of probabilistic and possibilistic data management \cite{Barbara92,Lee92,Pittarelli94,Dey96,Bosc96,Lakshmanan97,Fuhr97}. More recently, there have been successful initiatives to build working systems for the efficient execution of queries over uncertain data \cite{Boulos05,Cheng05,Widom05,Agrawal06,Sarma06,Re07, Huang09}.

In addition to specific works on uncertain data models and systems, this article builds over the large literature on relational join algorithms which is today consolidated and can be found in any text book on relational database management system architectures, e.g., \cite{Garcia-Molina08}. With regard to traditional join algorithms, we have reported and exemplified the few concepts necessary to understand the remaining of the paper in the introduction.

On the contrary, the problem of direct optimization of uncertain data is more recent and to the best of our knowledge the first work suggesting the usage of specific statistics on the uncertainty of the input relations is \cite{Sarma08}. Other works have also dealt with query optimization on probabilistic data without focusing on join algorithms \cite{Dalvi04,Re07,Antova08}. 

Given the importance of the join operator a number of probabilistic join algorithms have been proposed in the literature, and are complementary to our work because they deal with different aspects of this problem, as follows.

Probabilistic joins are useful when objects may match \emph{up to a certain degree}, differently from our work where we compute exact joins and the additional workload depends on our ignorance of the real values we are manipulating. These approximate probabilistic joins have been studied in
\cite{Kriegel06}, dealing with joins between similar objects, and \cite{Kriegel07}, focusing on nearest-neighbor joins.

Other works have studied the execution of probabilistic joins on streaming data \cite{Xiang09,Xiang10}, focusing on the specific constraints of this context.
Another situation where we often have uncertain data is that of spatial databases, where we may not know with certainty the shape of the objects.
 \cite{Ljosa08} studied how to join this kind of data, and here the underlying (spatial) uncertainty model is different with respect to the one adopted in our work.

Finally, in this article we have studied methods to reduce the number of comparisons between certain or uncertain tuples. When we have numerical values like probabilities attached to our uncertain tuples we may also use these values (confidences) to exclude some pairs of tuples or to return first those pairs maximizing their joint confidence. This topic is of particular interest and complexity within the framework of continuous uncertainty models, and has been studied with regard to indexes for uncertain data and also with regard to the join operator in \cite{Agrawal09join}. In this case we often speak of Threshold Probabilistic Join Queries. Preliminary main memory versions of some algorithms presented in this work were also sketched in \cite{MagnaniSUM08} together with an experimental testing on two example data sets.

For a more extensive comparative discussion of some of these algorithms the interested reader may consult a recent survey on this topic \cite{Kriegel09}.

%\item LOSSY APPROACHES - Trio, difference with loss-less join:
%\item Computation of probabilities - lazy approaches and MystiQ
%
%This approach has many pros, and some typical limitations of layered approaches. Using an underlying query engine we have several benefits in term of architectural simplicity. In addition, we can take advantage of the underlying optimization capabilities of the relational engine. At the same time, the underlying relational system \emph{is not aware} of the external data model, and it may not be able to perform optimizations and algorithms that are specific for that data model. This claim is supported by recent research efforts to define special-purpose indexes and statistics for uncertain data 
%\cite{Sarma08}.

%%%%%%%%%%%%%%%%%%%%%%%%%%%%%%%%%%%%%%%%%%%%%%%%%%%%%%%%%%%%%%%%%%%%%%%%%%%%%%%%%%%%%%%%%%%%%%%%%%%%%%%%%%%%%%%%%%%%%%
%%%%%%%%%%%%%%%%%%%%%%%%%%%%%%%%%%%%%%%%%%%%%%%%%%%%%%%%%%%%%%%%%%%%%%%%%%%%%%%%%%%%%%%%%%%%%%%%%%%%%%%%%%%%%%%%%%%%%%
%%%%%%%%%%%%%%%%%%%%%%%%%%%%%%%%%%%%%%%%%%%%%%%%%%%%%%%%%%%%%%%%%%%%%%%%%%%%%%%%%%%%%%%%%%%%%%%%%%%%%%%%%%%%%%%%%%%%%%
\section{Join algorithms}\label{algo}

In this section we introduce the algorithms object of this paper, with a preliminary discussion of their main characteristics. Code listings are in pl/pgSQL\footnote{http://www.postgresql.org/docs/8.4/static/plpgsql.html} and should be understandable with some basic familiarity with SQL and the usage of cursors. For simplicity we will assume that joins are performed on a single attribute (called \texttt{val} in the code). All the approaches are exemplified on a self-join of the table illustrated in Figure~\ref{fig:forbes-unc}.

\subsection{Nested loop and base join}

The nested loop approach is not different from the one presented in the introduction: it compares all uncertain tuples in the external relation with all uncertain tuples in the internal relation and the order of comparisons is the same of the nested loop example presented in the introduction  --- there are however more matches due to the additional values introduced in the uncertain version of our working example (Figure~\ref{fig:forbes-unc}), as illustrated in Figure~\ref{fig:unc_nestedloop}. 
The only specificity of the version for uncertain data is that an intersection operator (\&\&) is used to find matching values instead of a simple equality comparison (line 11):

\begin{lstlisting}
CREATE FUNCTION nestedloopjoin(tab1, tab2)
RETURNS TABLE AS
   OPEN cur1 FOR SELECT * FROM tab1;
   OPEN cur2 FOR SELECT * FROM tab2;
   LOOP
      FETCH NEXT FROM cur1 into rec1;
      EXIT WHEN rec1 IS NULL;
      LOOP
         FETCH NEXT FROM cur2 into rec2;
         EXIT WHEN rec2 IS NULL;
         IF rec1.val && rec2.val 
            RETURN NEXT concat(rec1, rec2);
         END IF;
         END LOOP;
      MOVE ABSOLUTE 0 FROM cur2;
      END LOOP;
   RETURN;
\end{lstlisting}

\begin{figure}
\begin{center}
\begin{tabular}{r||c|c|c|c|c|c|}
  & ut1 & ut2 & ut3 & ut4 & ut5 & ut6 \\
\hline \hline
ut1 & \textbf{1}  & \textbf{2} & 3  & 4 & 5 & 6 \\
\hline
ut2 & \textbf{7}  & \textbf{8}  & 9  & 10  & 11  & 12  \\ 
\hline
ut3 & 13  & 14 & \textbf{15} & 16 & 17  & \textbf{18}  \\ 
\hline
ut4 & 19  & 20 & 21  & \textbf{22}  &  23 & \textbf{24} \\ 
\hline
ut5 & 25  & 26  & 27 & 28  & \textbf{29} & 30\\ 
\hline
ut6 & 31  & 32  & \textbf{33}  & \textbf{34}  & 35 & \textbf{36}\\ 
\hline
\end{tabular} \hspace{.1cm}
\end{center}
\caption{Comparisons performed using a nested loop join on our uncertain working example. Numbers indicate the order of comparisons and boldface the matches found}
\label{fig:unc_nestedloop}
\end{figure}

In addition to our implementation of the uncertain nested loop algorithm we will also let the system execute directly the query:
\begin{tabbing}
SELECT * FROM tab1 JOIN tab2 ON tab1.val \&\& tab2.val
\end{tabbing}
to compare our results with the one obtained using the system relational query optimizer, i.e., without explicit support for uncertain data joins.
This will be called base join in the following experiments and corresponds to the following code:

\begin{lstlisting}
CREATE FUNCTION basejoin(tab1, tab2)
RETURNS TABLE AS
   OPEN cur FOR SELECT * FROM tab1
      JOIN tab2 ON tab1.val && tab2.val;    
   LOOP
      FETCH NEXT FROM cur into res;
      EXIT WHEN res IS NULL;
      RETURN NEXT res;
      END LOOP;
   RETURN;
\end{lstlisting}

%%%%%%%%%%%%%%%%%%%%%%%%%%%%%%%%%%%%%%%%%%%%%%%%%%%%%%%%%%%%%%%%%%%%%%%%%%%%%%%%%%%%%%%%%%%%%%%%%%%%%%%%%%%%%%%%%%%%%%
%%%%%%%%%%%%%%%%%%%%%%%%%%%%%%%%%%%%%%%%%%%%%%%%%%%%%%%%%%%%%%%%%%%%%%%%%%%%%%%%%%%%%%%%%%%%%%%%%%%%%%%%%%%%%%%%%%%%%%
\subsection{Sort join}

Our extension of the traditional sort join algorithm considers each uncertain value as a range, from its lower alternative value to the upper one, and every record is extended with two attributes (\texttt{lb} and \texttt{ub}) representing the bounds of the range. For example, the tuple ut1 in Figure~\ref{fig:forbes-unc} has ut1.lb=40 and ut1.ub=58. In this way, two tuples potentially match only if their ranges have a non-empty intersection. The code of the algorithm is the following:

\begin{lstlisting}
CREATE FUNCTION sortjoin(tab1, tab2)
RETURNS TABLE AS
   offset = 0;
   OPEN cur1 FOR SELECT * FROM tab1
      order by lb, ub;
   OPEN cur2 FOR SELECT * FROM tab2
      order by lb, ub;    
   LOOP
      FETCH NEXT FROM cur1 into rec1;
      EXIT WHEN rec1 IS NULL;
      MOVE -offset FROM cur2;
      offset = 0;
      LOOP
         FETCH NEXT FROM cur2 into rec2;
         offset = offset + 1;
         EXIT WHEN rec2 IS NULL;
         EXIT WHEN rec1.ub < rec2.lb;
         IF rec2.ub < rec1.lb AND offset = 1 
            offset = 0;
         END IF;
         IF rec2.ub >= rec1.lb AND
            rec1.val && rec2.val 
            RETURN NEXT concat(rec1, rec2);
         END IF;
         END LOOP;
      END LOOP;
   RETURN;
\end{lstlisting}

First uncertain tables are sorted by the lower bounds of the uncertain attributes (\texttt{lb}, lines 4-5 and 6-7).
Then we compare every record in the external relation (fetched at line 9) with its potential matches in the internal one (fetched at line 14). Let us consider our working example: we compare ut6 (having range [14,18]) against ut6, then with ut3 ([14,22]), ut4 ([18,19]) and ut5 ([\textbf{23},28]). At this point thanks to the ordering we already know that ut6 will not match neither ut2 nor ut1, because the upper value in ut6 is 18 and this is less than 23.  This condition is verified at line 17 and guarantees that no subsequent uncertain tuples in the (ordered) internal relation will match the uncertain tuple under consideration, as it happens with traditional sort joins over certain relations. This is then repeated for all tuples in the external input relation.

\begin{figure}
\begin{center}
\begin{tabular}{r||c|c|c|c|c|c|}
  & ut6 & ut3 & ut4 & ut5 & ut2 & ut1 \\
\hline \hline
ut6 & \textbf{1}  & \textbf{2} & \textbf{3}  & 4 &  &  \\
\hline
ut3 & \textbf{5}  & \textbf{6}  & 7  & 8  &   &   \\ 
\hline
ut4 & \textbf{9}  &  10 & \textbf{11}  &  12 &   &   \\ 
\hline
ut5 & 13  & 14  & 15  & \textbf{16}  & 17  &   \\ 
\hline
ut2 &   &   &   & 18 & \textbf{19}  & \textbf{20}  \\ 
\hline
ut1 &   &   &   &   & \textbf{21}  &  \textbf{22} \\ 
\hline
\end{tabular} \hspace{.1cm}
\end{center}
\caption{Comparisons performed using an extended sort join. Numbers indicate the order and boldface the matches --- notice that the input tuples have been sorted on the lower bounds of the join attribute}
\label{fig:sortjoin}
\end{figure}

It is worth noticing that this approach reduces to a traditional sort join when the join attribute is not uncertain, or when it contains a single option. Intuitively the difference in performance between this approach and a traditional sort join will be small when the uncertainty does not change significantly the join attribute, that is when the range of values remains close to a single point, while very large ranges may result in many unnecessary comparisons --- this intuition will be verified in the experimental testing. In Figure~\ref{fig:sortjoin} we have illustrated the order of comparisons and the found matches applying this extended algorithm to our working example. 

%%%%%%%%%%%%%%%%%%%%%%%%%%%%%%%%%%%%%%%%%%%%%%%%%%%%%%%%%%%%%%%%%%%%%%%%%%%%%%%%%%%%%%%%%%%%%%%%%%%%%%%%%%%%%%%%%%%%%%
%%%%%%%%%%%%%%%%%%%%%%%%%%%%%%%%%%%%%%%%%%%%%%%%%%%%%%%%%%%%%%%%%%%%%%%%%%%%%%%%%%%%%%%%%%%%%%%%%%%%%%%%%%%%%%%%%%%%%%
\subsection{Tuple-based join}

This approach transforms the input uncertain relations into traditional relations and joins them using existing algorithms. Then a postprocessing phase is required to re-build an uncertain relation. The main advantage of this approach is that joins are performed on traditional relations, therefore they depend only on the values contained in the input relations but not on how uncertainty is distributed. The price to pay comes from pre- and post-processing phases, together with the size and cardinality of the intermediate relations. Therefore, intuitively this approach can be used when the specific distribution of uncertain values makes direct methods like the extended sort join algorithm too complex.

To use this approach we need each tuple to be identified uniquely by an attribute that we call \texttt{xid} and that in our working example corresponds to the primary key of the relation.

\begin{lstlisting}
CREATE FUNCTION tuplejoin1(tab1, tab2)
RETURNS TABLE AS
   OPEN cur FOR SELECT tab1.*, tab2.*
      FROM flatjoin1(tab1, tab2) idxt
      join tab1 ON idxt.xid1 = tab1.xid
      join tab2 ON idxt.xid2 = tab2.xid;
   LOOP
      FETCH NEXT FROM cur into res;
      EXIT WHEN res IS NULL;
      RETURN NEXT res;
      END LOOP;
   RETURN;
\end{lstlisting}

\begin{figure}
\begin{center}
\begin{tabular}{|c|c|}
\hline
xid & \textbf{Net Worth} \\
\hline \hline
ut3 & 14 \\
\hline
ut6 & 14 \\
\hline
ut6 & 16 \\
\hline
ut3 & 16 \\
\hline
ut4 & 18 \\
\hline
ut6 & 18 \\
\hline
ut4 & 19 \\
\hline
ut3 & 22 \\
\hline
ut5 & 23 \\
\hline
ut5 & 25 \\
\hline
ut5 & 28 \\
\hline
ut2 & 37 \\
\hline
ut2 & 40 \\
\hline
ut1 & 40 \\
\hline
ut2 & 42 \\
\hline
ut2 & 47 \\
\hline
ut1 & 50 \\
\hline
ut1 & 53 \\
\hline
ut1 & 58 \\
\hline
\end{tabular}
% \hspace{.1cm}
%\begin{tabular}{|c|c|}
%\hline
%xid & \textbf{Net Worth} \\
%\hline \hline
%ut3 & 14 \\
%\hline
%ut6 & 14 \\
%\hline
%ut6 & 16 \\
%\hline
%ut3 & 16 \\
%\hline
%ut4 & 18 \\
%\hline
%ut6 & 18 \\
%\hline
%ut4 & 19 \\
%\hline
%ut3 & 22 \\
%\hline
%ut5 & 23 \\
%\hline
%ut5 & 25 \\
%\hline
%ut5 & 28 \\
%\hline
%ut2 & 37 \\
%\hline
%ut2 & 40 \\
%\hline
%ut1 & 40 \\
%\hline
%ut2 & 42 \\
%\hline
%ut2 & 47 \\
%\hline
%ut1 & 50 \\
%\hline
%ut1 & 53 \\
%\hline
%ut1 & 58 \\
%\hline
%\end{tabular}
\end{center}
\caption{Flattening of our working uncertain table used to compute \texttt{flatjoin1} (notice that the same table is used as both external and internal relation in this example, as we are computing a self-join)}
\label{fig:tuple-ex1}
\end{figure}

\begin{figure}
\begin{center}
\begin{tabular}{|c|c|c|c|}
\hline
xid & \textbf{Net Worth} & xid & \textbf{Net Worth} \\
\hline \hline
ut3 & 14 & ut3 & 14\\
\hline
ut3 & 14 & ut6 & 14 \\
\hline
ut6 & 14 & ut3 & 14 \\
\hline
ut6 & 14 & ut6 & 14 \\
\hline
ut6 & 16 & ut6 & 16 \\
\hline
ut6 & 16 & ut3 & 16 \\
\hline
ut3 & 16 & ut6 & 16 \\
\hline
ut3 & 16 & ut3 & 16 \\
\hline
ut4 & 18 & ut4 & 18 \\
\hline
ut4 & 18 & ut6 & 18 \\
\hline
ut6 & 18 & ut4 & 18 \\
\hline
ut6 & 18 & ut6 & 18 \\
\hline
ut4 & 19 & ut4 & 19 \\
\hline
ut3 & 22 & ut3 & 22 \\
\hline
ut5 & 23 & ut5 & 23 \\
\hline
ut5 & 25 & ut5 & 25 \\
\hline
ut5 & 28 & ut5 & 28 \\
\hline
ut2 & 37 & ut2 & 37 \\
\hline
ut2 & 40 & ut2 & 40 \\
\hline
ut2 & 40 & ut1 & 40 \\
\hline
ut1 & 40 & ut2 & 40 \\
\hline
ut1 & 40 & ut1 & 40 \\
\hline
ut2 & 42 & ut2 & 42 \\
\hline
ut2 & 47 & ut2 & 47 \\
\hline
ut1 & 50 & ut1 & 50 \\
\hline
ut1 & 53 & ut1 & 53 \\
\hline
ut1 & 58 & ut1 & 58 \\
\hline
\end{tabular} \hspace{.1cm}
\begin{tabular}{|c|c|}
\hline
xid & xid \\
\hline \hline
ut3 & ut3 \\
\hline
ut3 & ut6 \\
\hline
ut6 & ut3 \\
\hline
ut6 & ut6 \\
\hline
ut4 & ut4 \\
\hline
ut4 & ut6 \\
\hline
ut6 & ut4 \\
\hline
ut6 & ut6 \\
\hline
ut5 & ut5 \\
\hline
ut2 & ut2 \\
\hline
ut2 & ut1 \\
\hline
ut1 & ut2 \\
\hline
ut1 & ut1 \\
\hline
\end{tabular}
\end{center}
\caption{Partial and final result (with all distinct pairs of ids of matching uncertain tuples) of the join between the flattening of the input tables, line 4 in the code}
\label{fig:tuple-ex2}
\end{figure}

The first part of the algorithm consists in the flattening of the input relations, whose result is illustrated in  Figure~\ref{fig:tuple-ex1}. Having all the alternative values of the uncertain attribute and the corresponding record identifiers we can thus find matching records using a traditional join. The result of this phase is illustrated in Figure~\ref{fig:tuple-ex2}, and this first part of the algorithm corresponds to the \texttt{flatjoin1} function at line 4 of the code. Finally, the original uncertain tuples are recovered by joining the obtained pairs of identifiers with the two input uncertain relations, to recollect all the other attributes.

In summary, this approach consists in building traditional relational tables, join them using traditional algorithms and get back all the uncertain tuples with matching values. This general approach can be implemented in different ways. First, it is interesting to notice that in the Trio system relations are already stored in this format, and the flattening phase would not be necessary. In addition, notice that in the flattening phase exemplified in Figure~\ref{fig:tuple-ex1} we may decide not to project only on the \textbf{xid} and \textbf{Net Worth} attributes, but to keep also all the other attributes from the input tables. As a result, we will not need the subsequent joins (lines 5--6) to recollect the information projected out in the previous approach. However, in this way we have to work with larger intermediate tables replicating all the other attributes for each alternative value of the join attribute. In the following experiments we will evaluate also this variation, that will be indicated as \emph{tuple join 2}.

\begin{lstlisting}
CREATE FUNCTION tuplejoin2(tab1, tab2)
RETURNS TABLE AS
   OPEN cur FOR SELECT distinct tab1.*, tab2.*
      FROM flatjoin2(tab1, tab2)
   LOOP
      FETCH NEXT FROM cur into res;
      EXIT WHEN res IS NULL;
      RETURN NEXT res;
      END LOOP;
   RETURN;
\end{lstlisting}

Notice that we now perform only one join, indicated at line 4 (\texttt{flatjoin2}). This is computed on flattened tables like the one represented in Figure~\ref{fig:forbes-unc-flat} with regard to our working example. Intuitively, this version of the tuple join approach has the advantage of having to compute only one join but in general it has to deal with larger intermediate relations keeping all the data from the input tables, so that it does not have to retrieve it later.

\begin{figure}
\begin{center}
\begin{tabular}{|c||l|l|l|l|l|l|}
\hline
xid & \textbf{Name} & \textbf{Surname} & \textbf{Net Worth} & \textbf{Age} & \textbf{val} \\
\hline \hline
ut1 & William H. & Gates III & \{53,50,40,58\} & 54 & 53 \\
\hline
ut1 & William H. & Gates III & \{53,50,40,58\} & 54 & 50\\
\hline
ut1 & William H. & Gates III & \{53,50,40,58\} & 54 & 40\\
\hline
ut1 & William H. & Gates III & \{53,50,40,58\} & 54 & 58\\
\hline
ut2 & Warren & Buffett & \{47,40,37,42\} & 79 & 47\\
\hline
ut2 & Warren & Buffett & \{47,40,37,42\} & 79 & 40\\
\hline
ut2 & Warren & Buffett & \{47,40,37,42\} & 79 & 37\\
\hline
ut2 & Warren & Buffett & \{47,40,37,42\} & 79 & 42\\
\hline
ut3 & Paul & Allen & \{14,16,22\} & 57 & 14\\
\hline
ut3 & Paul & Allen & \{14,16,22\} & 57 & 16\\
\hline
ut3 & Paul & Allen & \{14,16,22\} & 57 & 22\\
\hline
ut4 & Lawrence & Page & \{18,19\} & 37 & 18\\
\hline
ut4 & Lawrence & Page & \{18,19\} & 37 & 19\\
\hline
ut5 & Lawrence & Ellison & \{28,23,25\} & 65 & 28\\
\hline
ut5 & Lawrence & Ellison & \{28,23,25\} & 65 & 23\\
\hline
ut5 & Lawrence & Ellison & \{28,23,25\} & 65 & 25\\
\hline
ut6 & Michael & Dell & \{14,16,18\} & 45 & 14\\
\hline
ut6 & Michael & Dell & \{14,16,18\} & 45 & 16\\
\hline
ut6 & Michael & Dell & \{14,16,18\} & 45 & 18\\
\hline
\end{tabular} \hspace{.1cm}
\end{center}
\caption{Flattening of our working example used to compute \texttt{flatjoin2}}
\label{fig:forbes-unc-flat}
\end{figure}

%%%%%%%%%%%%%%%%%%%%%%%%%%%%%%%%%%%%%%%%%%%%%%%%%%%%%%%%%%%%%%%%%%%%%%%%%%%%%%%%%%%%%%%%%%%%%%%%%%%%%%%%%%%%%%%%%%%%%%
%%%%%%%%%%%%%%%%%%%%%%%%%%%%%%%%%%%%%%%%%%%%%%%%%%%%%%%%%%%%%%%%%%%%%%%%%%%%%%%%%%%%%%%%%%%%%%%%%%%%%%%%%%%%%%%%%%%%%%
\subsection{Index join}

We conclude this section with a variation of a traditional index join. Here every record \texttt{ut} is augmented with an attribute (called \texttt{b} in the following code) representing a segment with extreme points ut.lb and ut.ub. In addition we use a spatial index on this attribute, so that when we want to look at potential matches we access the index and extract all records with an intersecting segment:\\

\begin{lstlisting}
CREATE FUNCTION indexjoin(tab1, tab2)
RETURNS TABLE AS
   OPEN cur FOR SELECT * FROM tab1
      JOIN tab2 ON tab1.b && tab2.b
      WHERE tab1.val && tab2.val;    
   LOOP
      FETCH NEXT FROM cur into res;
      EXIT WHEN res IS NULL;
      RETURN NEXT res;
   END LOOP;
   RETURN;
\end{lstlisting}
	    
The effect of this code is to execute a query very similar to the one of the base join algorithm (lines 3-5):
\begin{tabbing}
SELECT *\\
FROM tab1 JOIN tab2 ON tab1.\textbf{b} \&\& tab2.\textbf{b}\\
WHERE tab1.val \&\& tab2.val
\end{tabbing}
The difference consists in the fact that the join is now performed on the indexed attributes. In this way the relational optimizer can use the index to identify potentially matching records, i.e., with a non empty intersection on the \texttt{b} attribute. All these are then checked against the WHERE predicate, selecting only those with a real match.

\section{Experimental results}\label{exp}

The algorithms presented in the previous section have been tested on several synthetic datasets and a real uncertain dataset obtained by extending the DBLP database with information about author institutions.

Synthetic uncertain tables contain a primary key, an uncertain join attribute and the additional attributes needed by the algorithms as previously explained (lower bound, upper bound - for sort join - and indexed uncertainty interval - for index join).
An example of these data is presented in Table~\ref{example-syn}.

To easily understand the following experiments we can use this table to do some examples of the terminology adopted in the remaining of the section. We call \emph{cardinality of an uncertain table} the number of uncertain records, four in the example (t0, t1, t2, t3). We call \emph{cardinality of an uncertain tuple} the number of alternative values contained in it, three for each tuple in the example (224.480, 224.482 and 224.484 for the first tuple). In the following experiments it is also important to consider the width of the interval spanned by these alternative values ([224.480, 224.484] for the first tuple in the example):  when values of one uncertain tuple are spread over a larger interval the number of records in the other input relation potentially matching it will typically increase. We will thus indicate with \emph{spreading} the number of potentially matching records. Another tested parameter is the \emph{percentage} of uncertain tuples, which is $100\%$ in our example. The DBLP dataset with the uncertain affiliations of the authors is described in \cite{Hideaki10}.

In this section we describe the experimental settings and the obtained results, and in the next section we provide an interpretation of these results. All the experiments have been conducted on a workstation with Linux Ubuntu 2.6.32.4, a 2GHz CPU, a Toshiba MK2555GS ATA disk and 3GB RAM. To be sure that the results are not influenced by the actual status of the buffers we restart the DBMS after every join and empty main memory using the Linux instruction \texttt{sync; echo 3 | sudo tee /proc/sys/vm/drop\_caches}. The DBMS used in the experiment is PostgreSQL 8.4, and statistics on disk accesses are obtained using the \texttt{iostat} shell command. The sequence of tests, system restarting, memory cleaning and time and I/O statistics gathering has been automated using a Java program connecting to the database through JDBC.

\begin{table*}[htdp]
\caption{Structure of the synthetic tables used in the experiments with some example records}\begin{center}
\begin{tabular}{|l|l|l|l|l|}
\hline
\textbf{pk} & \textbf{lb} & \textbf{ub} & \textbf{b} & \textbf{val} \\
\hline
\hline
t0 & 224.480 & 224.484 & ((224.480,0), (224.484,0)) & \{224.480, 224.482, 224.484 \} \\
\hline
t1 & 954.463 & 954.467 & ((954.463,0), (954.467,0)) & \{954.463, 954.465, 954.467 \} \\
\hline
t2 & 133.374 & 133.378 & ((133.374,0), (133.378,0)) & \{133.374,133.376,133.378 \} \\
\hline
t3 & 12.000 & 12.004 & ((12.000,0), (12.004,0)) & \{12.000, 12.002, 12.004 \} \\
\hline
\end{tabular}
\end{center}
\label{default}
\label{example-syn}
\end{table*}
%
%
%The following parameters are used in the experiments to evaluate the efficiency of the algorithms:
%\begin{itemize}
%\item \textbf{Size}: number of uncertain tuples in each input relation. In the experiments varying the size of the relations we will use only uncertain tuples, each with two alternative values inside an interval of X units.
%\item \textbf{Percentage}: percentage of uncertain tuples in each input relation.
%\item \textbf{Uncertainty interval}: the interval inside with uncertain values are contained. For example, with an interval of 10 we may have an uncertain value like \{2010, 2013, 2020\}. 
%\item \textbf{Selectivity}: represents the percentage of tuples in the first relation matching tuples in the second. In the experiments evaluating the impact of selectivity on the efficiency of the algorithms all tuples will be uncertain.
%\end{itemize}

\subsection{Varying cardinality of input relations}

\begin{figure}
\begin{center}
\includegraphics[angle=-90,width=.5\columnwidth]{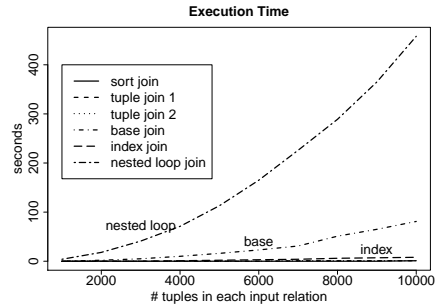}
\end{center}
\caption{All algorithms tested on small input relations (curves regarding sort and tuple joins are all overlapping on the lower curve with y=0, as these algorithms take less than one seconds on all test datasets)}
\label{size_small}
\end{figure}

\begin{figure}
\begin{center}
\includegraphics[angle=-90,width=.5\columnwidth]{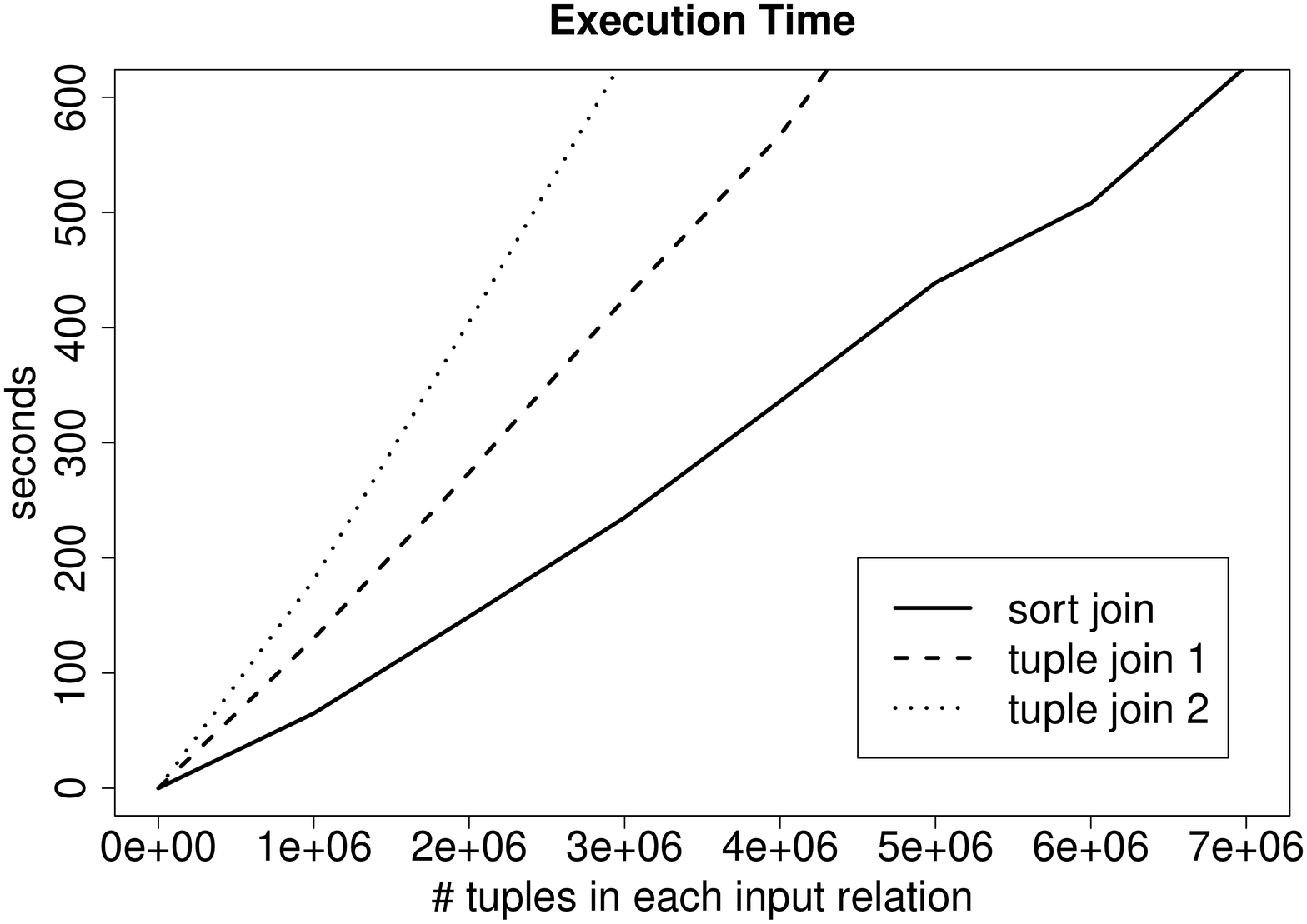}
\end{center}
\caption{Tuple joins and Sort join tested on large relations (time)}
\label{size}
\end{figure}

\begin{figure}
\begin{center}
\includegraphics[angle=-90,width=.5\columnwidth]{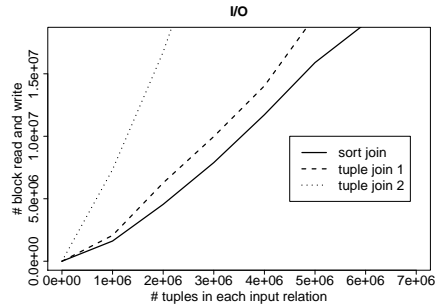}
\end{center}
\caption{Tuple joins and Sort join tested on large relations (IO)}
\label{size_io}
\end{figure}

In these experiments joins are performed between tables with a varying number of uncertain tuples and the following parameters:
\begin{itemize}
\item \textbf{Input table cardinality: } Varying from 1.000 (all algorithms) to 10.000.000 (only sort- and tuple-joins)
\item \textbf{Cardinality of result: } Same as input relations
\item \textbf{Cardinality of uncertain tuples: } 3
\item \textbf{Percentage of uncertain tuples: } 100\%
\item \textbf{Spreading: } 1
\end{itemize}
The results of the experiments are illustrated in Figures~\ref{size_small}, \ref{size}, \ref{size_io}.

In Figure~\ref{size_small} we show the execution time of all algorithms run on small relations. Here sort join and tuple join algorithms prove to be more efficient than the others, taking 0 to 1 second to perform every join task. Index-join is also very fast, while base join and nested loop join are slow even with these small datasets. Sort and Tuple joins are the only algorithms for which we could compute joins between tables with millions of uncertain tuples. In Figures~\ref{size} and \ref{size_io} we have shown the execution time and the number of disk accesses for these algorithms, varying the number of uncertain tuples in each input relation. If we compare the two plots we can see that time complexity is determined by the number of I/O operations --- when this will be the case for the following experiments we will only include the plots with execution times.

\subsection{Varying size of input relations, top results}

\begin{figure}
\begin{center}
\includegraphics[angle=-90,width=.5\columnwidth]{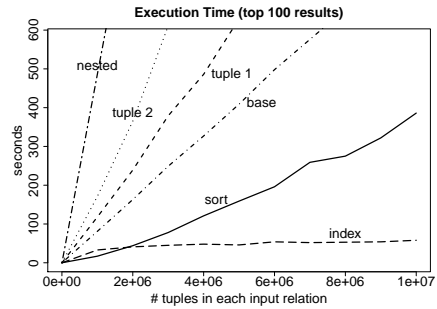}
\end{center}
\caption{Time to retrieve the first 100 tuples in the result}
\label{size_top100}
\end{figure}

\begin{figure}
\begin{center}
\includegraphics[angle=-90,width=.5\columnwidth]{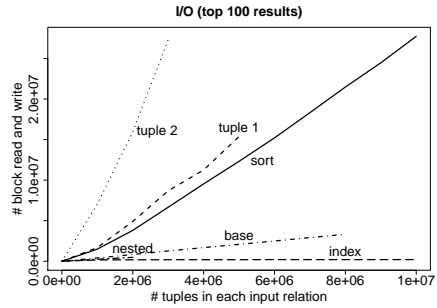}
\end{center}
\caption{Number of I/O operations to retrieve the first 100 tuples in the result}
\label{size_top100_io}
\end{figure}

When we perform joins using GUIs the typical behavior of modern database management systems consists in retrieving only the first few tuples of the result, usually with a default of about 100 tuples. Additional tuples are fetched into memory only if explicitly required by the user. As a consequence, it is interesting to evaluate how fast our algorithms are to start producing results and to compute a small subset of the resulting tuples. The results of these tests have been illustrated in Figures \ref{size_top100}, \ref{size_top100_io}.

First, notice that differently from the previous experiments all algorithms may deal with large tables with millions of uncertain tuples. Here, the algorithms in order of increasing efficiency are: nested loop, tuple (second and first version), base join, sort join, and index join, which is by far the most efficient approach.

Second, the performance of tuple-join does not improve significantly with respect to the previous tests.

Finally, looking at Figure~\ref{size_top100_io} we can notice that the relative efficiency of the algorithms with respect to execution time or number of I/O operations changes, showing that the execution time is not only determined by disk accesses.

\subsection{Varying spreading of alternative values}

\begin{figure}
\begin{center}
\includegraphics[angle=-90,width=.5\columnwidth]{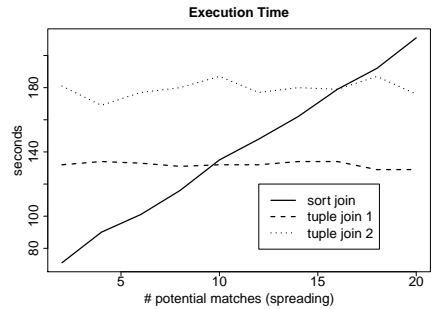}
\end{center}
\caption{Tuple-joins and Sort-join tested on relations whose uncertain tuples contain values spanning an increasingly large interval (time)}
\label{spre}
\end{figure}

\begin{figure}
\begin{center}
\includegraphics[angle=-90,width=.5\columnwidth]{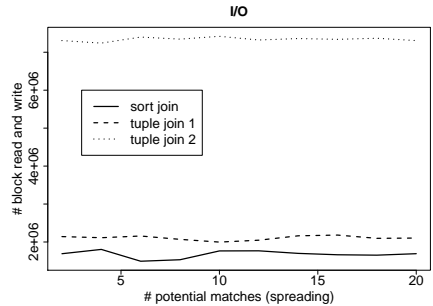}
\end{center}
\caption{Tuple-joins and Sort-join tested on relations whose uncertain tuples contain values spanning an increasingly large interval (IO)}
\label{spre_io}
\end{figure}

In the previous tests the alternative values inside each tuple were very close to the same value, e.g., \{10,12,14\}. In this way every uncertain tuple in one relation potentially matched only one tuple in the other relation. In this test we increase the interval spanned by the alternative values to increase the number of potential matches, indicated by the \emph{spreading} parameter.
\begin{itemize}
\item \textbf{Input table cardinality: } 1.000.000
\item \textbf{Cardinality of result: } 1.000.000
\item \textbf{Cardinality of uncertain tuples: } 3
\item \textbf{Percentage of uncertain tuples: } 100\%
\item \textbf{Spreading: } Varying from 1 to 20
\end{itemize}
The results of the test are presented in Figures \ref{spre}, \ref{spre_io}.

Figure \ref{spre} shows an interesting behavior: the performance of both tuple join approaches is constant, while the execution time of the sort join depends on the variation of the \emph{spreading} parameter.

Figure \ref{spre_io} shows the number of I/O operations needed by the three algorithms for the same experiment. In this case all curves are constant.

\subsection{Varying tuple cardinality and percentage of uncertain records}
\begin{figure}
\begin{center}
\includegraphics[angle=-90,width=.5\columnwidth]{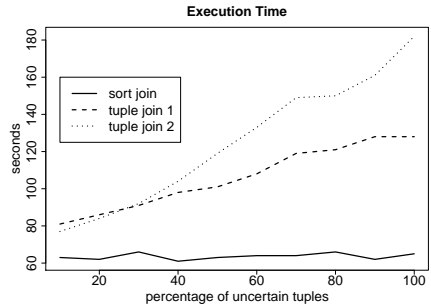}
\end{center}
\caption{Tuple-joins and Sort-join tested on relations with varying percentage of uncertain tuples}
\label{perc}
\end{figure}

\begin{figure}
\begin{center}
\includegraphics[angle=-90,width=.5\columnwidth]{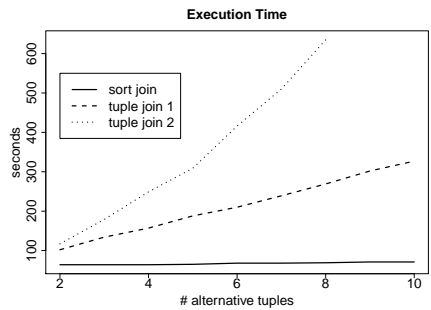}
\end{center}
\caption{Tuple-joins and Sort-join tested on relations whose uncertain tuples contain an increasing number of alternative values (time)}
\label{card}
\end{figure}

\begin{figure}
\begin{center}
\includegraphics[angle=-90,width=.5\columnwidth]{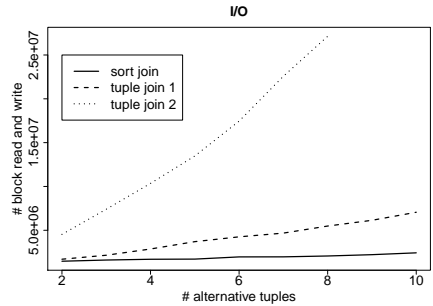}
\end{center}
\caption{Tuple-joins and Sort-join tested on relations whose uncertain tuples contain an increasing number of alternative values (IO)}
\label{card_io}
\end{figure}

In these experiments we vary the percentage of uncertain tuples in each input relation, and the cardinality of uncertain tuples, i.e., the number of alternative values contained in each tuple. The first set of tests corresponds to the following parameters:
\begin{itemize}
\item \textbf{Input table cardinality: } 1.000.000
\item \textbf{Cardinality of result: } 1.000.000
\item \textbf{Cardinality of uncertain tuples: } 3
\item \textbf{Percentage of uncertain tuples: } Varying from 10\% to 100\%
\item \textbf{Spreading: } 1
\end{itemize}
while the tests on the cardinality of the uncertain tuples have been performed with the following parameters:
\begin{itemize}
\item \textbf{Input table cardinality: } 1.000.000
\item \textbf{Cardinality of result: } 1.000.000
\item \textbf{Cardinality of uncertain tuples: } Varying from 2 to 10 
\item \textbf{Percentage of uncertain tuples: } 100\%
\item \textbf{Spreading: } 1
\end{itemize}
The results of the test are presented in Figures \ref{perc}, \ref{card}, \ref{card_io}.

Figures \ref{perc} and \ref{card} show an opposite behavior with regard to the previous experiments: now the complexity of tuple joins depends on the tested parameters, while the sort join has almost a constant behavior. Figure~\ref{card_io}, showing the number of I/O operations performed by the algorithms, allows us to appreciate a small dependency of the sort join on the number of alternative values without a perceivable effect on execution time with respect to the other algorithms.

%
%\begin{figure}
%\begin{center}
%\includegraphics[angle=-90,width=.5\columnwidth]{images/sele}
%\end{center}
%\caption{}
%\label{}
%\end{figure}
%
%\begin{figure}
%\begin{center}
%\includegraphics[angle=-90,width=.5\columnwidth]{images/sele_io}
%\end{center}
%\caption{}
%\label{}
%\end{figure}

\subsection{Query behavior on DBLP dataset}

We conclude the experimental analysis of our algorithms with a real dataset already used to evaluate indexes on uncertain data. While it is still difficult to find large and publicly available uncertain datasets because uncertain data management capabilities have not been incorporated into the mainstream relational database management systems, uncertain datasets are easily generated as a result of the integration of certain databases \cite{magnaniJDIQ10}. The data used in the following tests have been obtained by integrating DBLP author data with author affiliations, not present in the DBLP database and obtained via the Google API. This integration process is evidently uncertain, and up to ten possible alternative institutions have been recorded for each author. Additional details about these data can be found in \cite{Hideaki10}.

In this section we evaluate the following query:\\

\noindent\texttt{SELECT * FROM author A JOIN institution I\\on A.institution = I.id}\\\\
over two tables with respectively about 700.000 authors and about 6.000 institutions,
using all the algorithms. The resulting table contains about 2.600.000 records.

In this case the test shows three orders of efficiency: the two versions of the tuple join approach, taking a few hundred seconds, index and base join, taking a few thousand seconds, and sort and nested loop join, taking a few ten thousand seconds. In addition, it is interesting to evaluate the time needed to compute not only the whole result, but also smaller parts of the output.

In Figure~\ref{dblp tests} we have represented for each join its behavior in time, i.e., the number of records computed after $t$ seconds. Therefore, the slope of the curve represents the speed of producing results at time $t$. This will be clear by looking at the plots: both tuple join approaches do not output any result for a while (about 230 and 150 seconds respectively), then they start producing tuples very quickly until all the join has been computed. All the other approaches start computing tuples at once, with an almost linear behavior for index, sort and nested loop join and a decreasing speed for the base join, which is very fast at the beginning and slows down in time.

\section{Interpretation of the experimental results}
\label{interpretation}

\begin{figure}
\begin{center}
\includegraphics[angle=-90,width=.45\columnwidth]{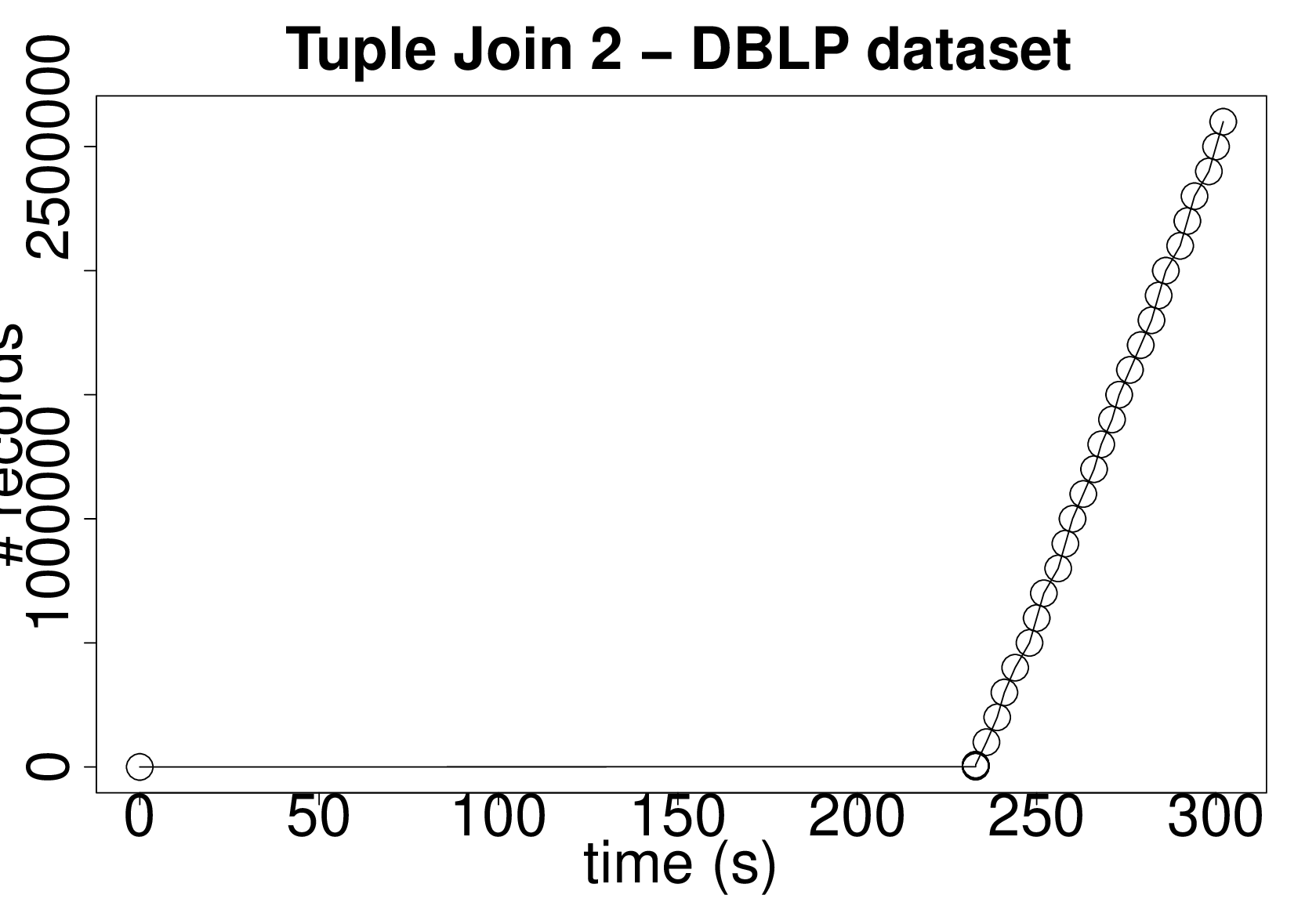}
\includegraphics[angle=-90,width=.45\columnwidth]{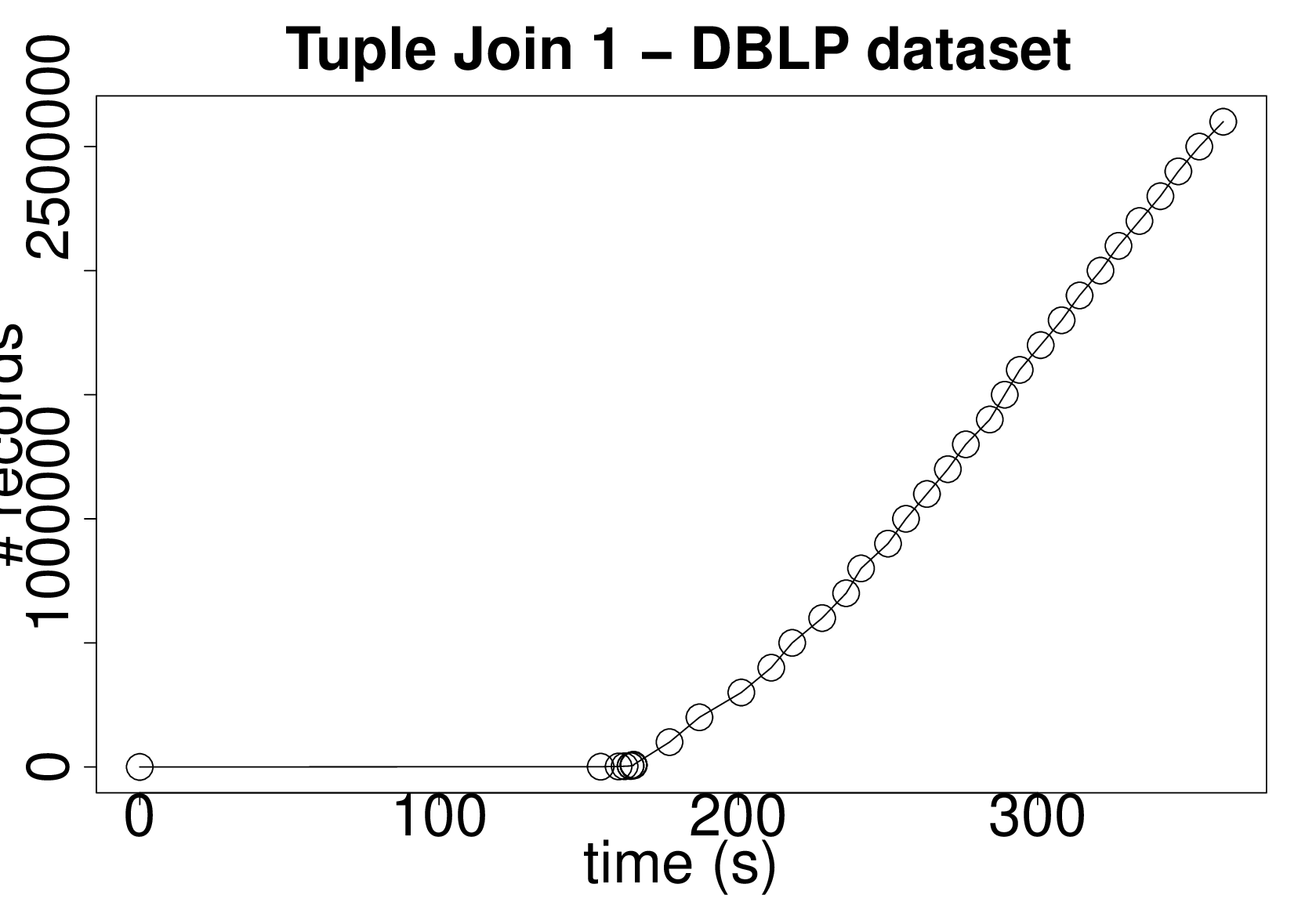}
\includegraphics[angle=-90,width=.45\columnwidth]{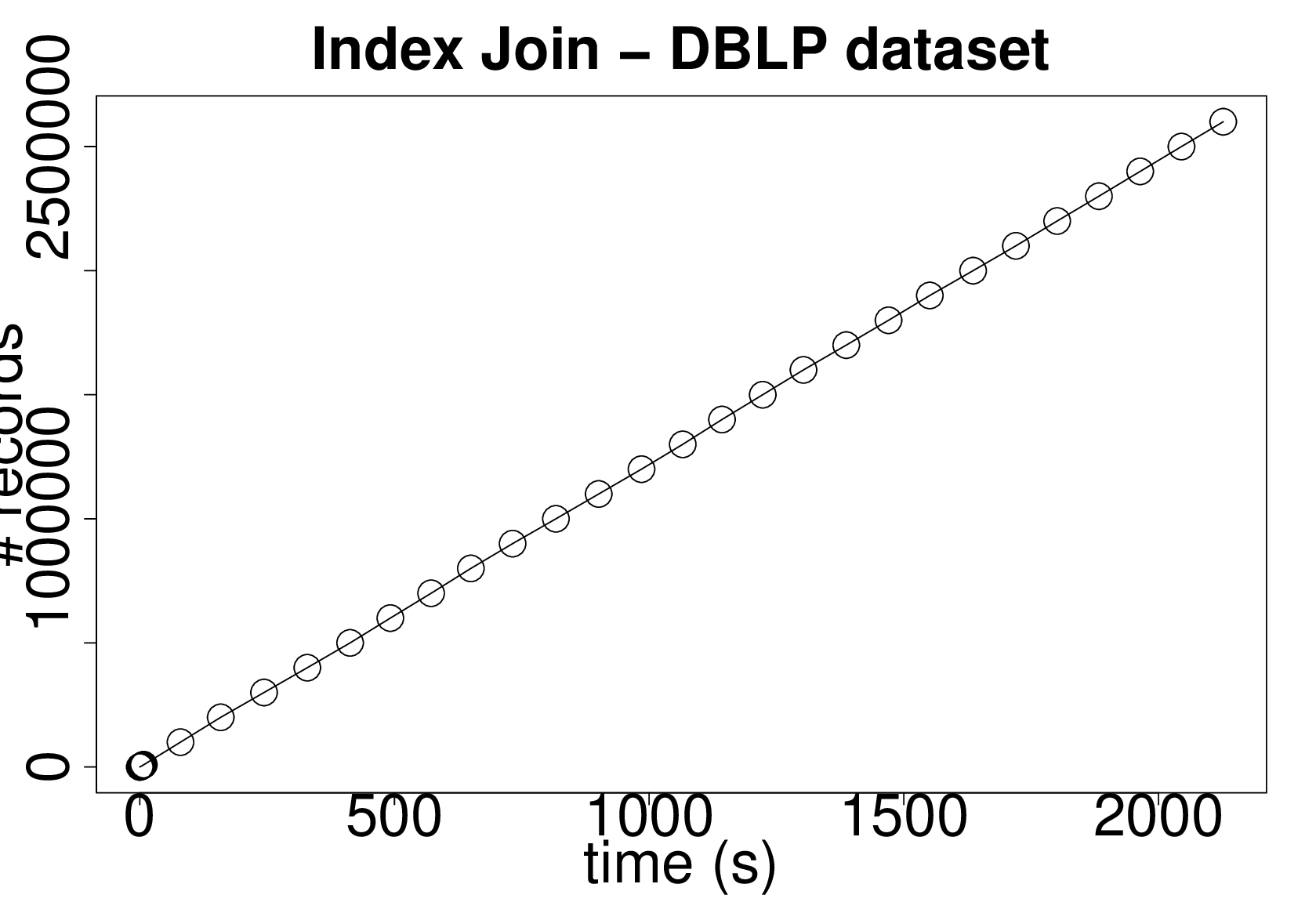}
\includegraphics[angle=-90,width=.45\columnwidth]{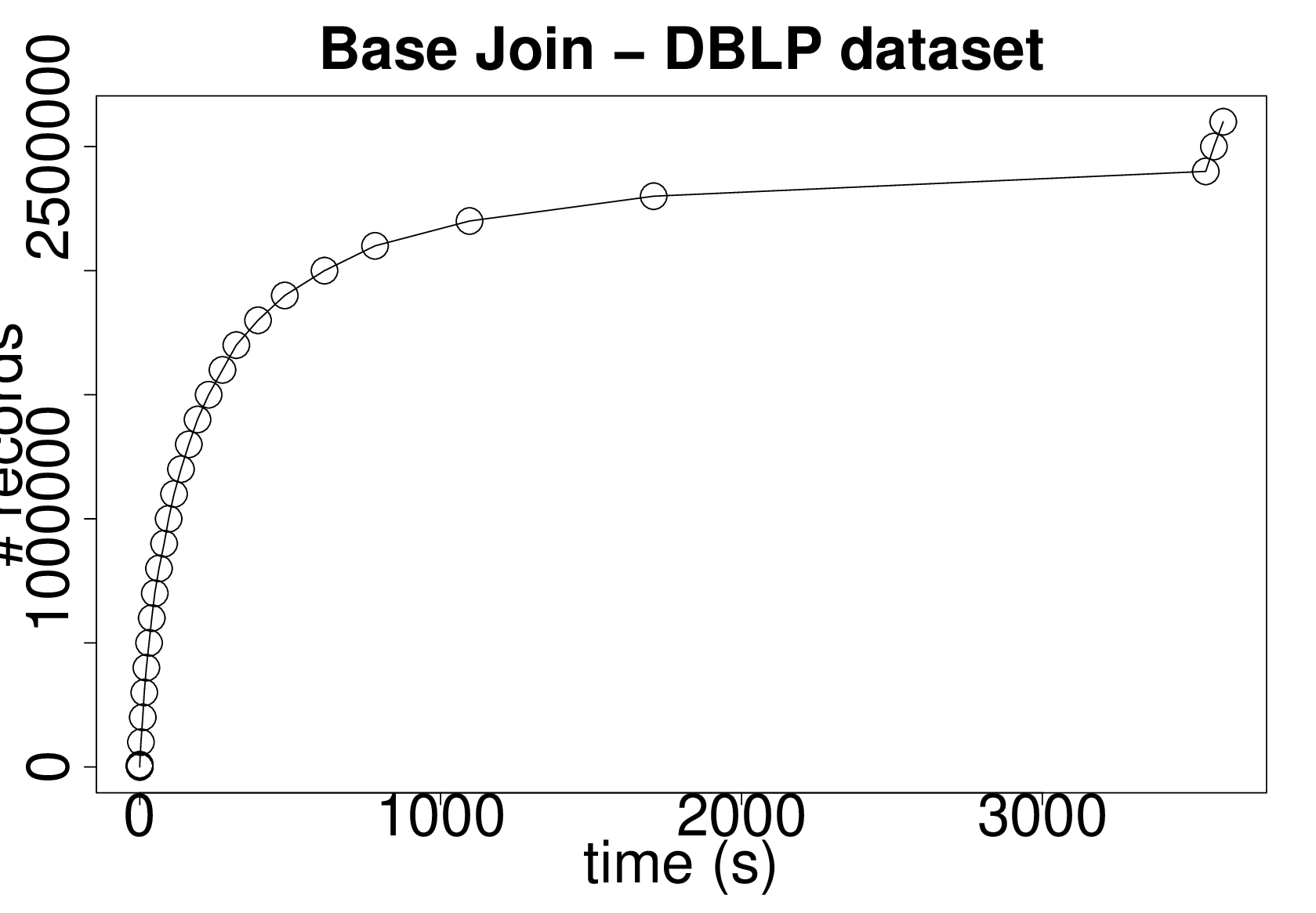}
\includegraphics[angle=-90,width=.45\columnwidth]{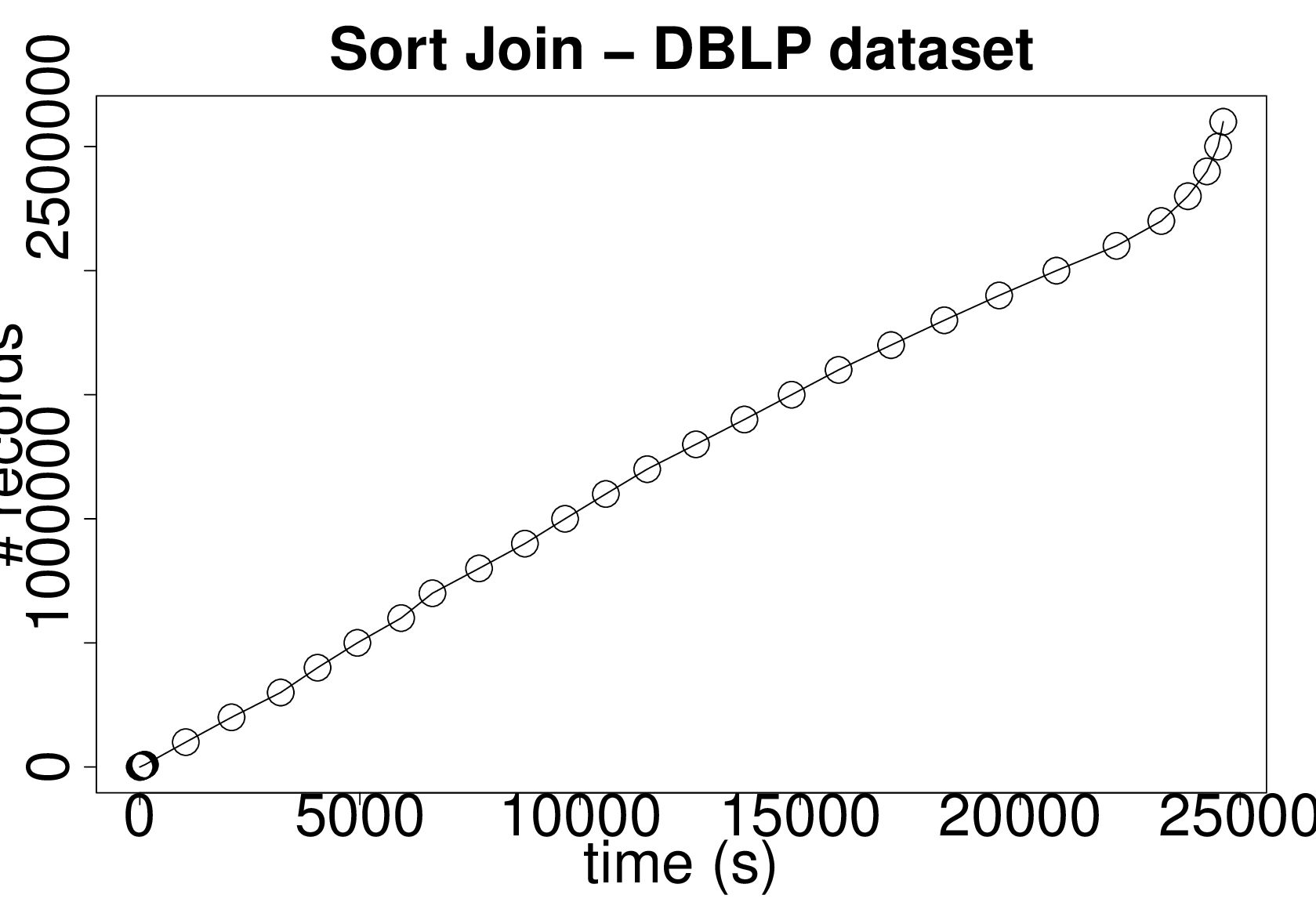}
\includegraphics[angle=-90,width=.45\columnwidth]{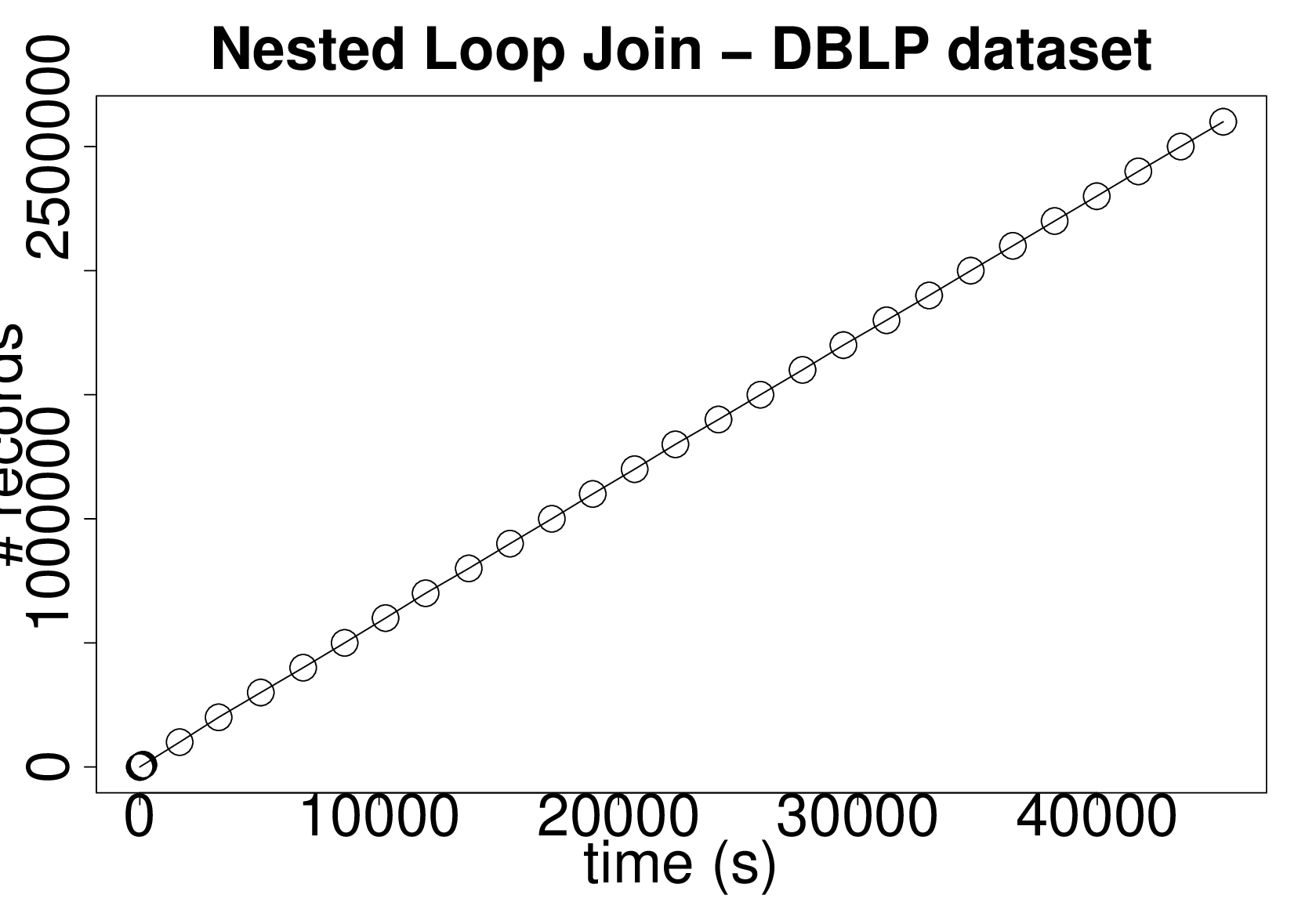}
\end{center}
\caption{Behavior of all join algorithms on the DBLP query.}
\label{dblp tests}
\end{figure}

In this section we provide our interpretations of the experimental results presented in Section~\ref{exp}.

The cardinality of the input relations obviously affects the performance of all algorithms. Among these, base join and nested-loop join are in general the slower methods (Figure~\ref{size_small}) because they do not provide any specific support for uncertain data (we remind the reader that the base join is the method chosen by the Postgres Query Optimizer). The index join is more efficient but still requires many disk accesses to retrieve the matching tuples.  Figure~\ref{size} shows that when alternative values are distributed in a small interval (where the meaning of \emph{small} is not absolute but depends on the number of matches with records in the other relation, as tested in following experiments) both sort join and tuple joins can deal with very large relations.

When it comes to compute only a few results, however, the index-based approach is the best one, because in our experimental setting it just requires one disk access for each tuple in the result, in addition to an index search (Figure~\ref{size_top100}). It is interesting to notice that the only algorithms that do not significantly improve their performance in this case are the tuple joins, because before starting to output the results they have to compute the complete join (tuple join 1) and perform a \texttt{distinct} operation (tuple join 2). Therefore, when they start producing the output they have already done most of their work.

Figures~\ref{size_io} and \ref{size_top100_io} show that the execution time of these tests is determined by the number of I/O operations for all approaches except base and nested loop join, where the execution time depends on the number of main memory operations.

Figures \ref{spre} and \ref{spre_io} show that the execution time of tuple-based approaches does not depend on the distribution of the alternative values inside each uncertain attribute. In fact, these methods split the alternative values into traditional relational tuples and thus perform a join between relations from which uncertainty has been temporarily removed. The sort join approach works directly on uncertain tuples, and is thus influenced by the distribution of uncertainty. In particular, when the interval inside which the alternative values are distributed increases this method has to perform a lot of potentially unnecessary comparisons, and its performance decreases --- in the worst case it behaves like a nested loop (or nested block loop) join, as it happens with traditional sort joins in traditional relational systems when many records have the same value on the join attribute in both input relations.

On the contrary, Figures \ref{perc} and \ref{card} highlight how the performance of the sort join depends on the cardinality of the uncertain relations, i.e., the number of uncertain tuples, while tuple based approaches depend on the number of alternative values. As an example, if an input relation contains 1.000.000 uncertain tuples with 10 alternative values each, the sort join will manipulate 1.000.000 uncertain tuples while the tuple join algorithm will operate on a traditional relation with 10.000.000 tuples. Therefore, when we add more alternative values or increase the percentage of tuples with multiple alternative values without spreading these values into a large interval, the performance of the sort join will be almost constant while tuple joins will be significantly affected.

Figure~\ref{card_io} shows that also the sort join method has a small dependency on the cardinality of the uncertain tuples: adding alternative values to an uncertain tuple increases the size of the input relations and thus requires additional I/O operations, as highlighted by the slight slope of the sort join curve.

Finally, Figure~\ref{dblp tests} provides some evidence regarding the behavior of these methods in time with respect to a single join, supporting and motivating some of the behaviors observed in the previous experiments. From the figure it is clear how both tuple join approaches have a preprocessing phase followed by a fast generation of the results. These tests also highlight how the sort join approach should not be used when alternative values are not distributed around specific values: in this case every author was associated to different institutions, and there is no reason why two institutions associated to the same author should be close to each other in the \texttt{Institution} table. As a consequence the behavior of this method is similar to the one of the nested loop join, though faster\footnote{The increase in speed at the end of the execution depends on the fact that only a few records remain to be tested at the end, therefore the number of main memory comparisons decreases}.

\section{Discussion and conclusion}

Besides other specific results, the experimental analysis described in Section~\ref{exp} highlights two main points. The first, as expected, is that joining uncertain relations with the methods introduced in the first part of this paper is in general more complex than joining certain relations. Depending on the specific algorithm, the increment in size of the input relations due to the additional values to store, the distribution of these values or the need to recompose uncertain tuples increase the computation time. The second point is that different algorithms are influenced by different factors. Therefore it is possible to choose an efficient algorithm depending on the input data.

More in detail, we have shown that tuple-based approaches have a time complexity which is independent of the distribution of uncertain values in the data, and corresponds to the execution of three traditional joins (or one join with duplicate removal) on larger certain relations. However, their performance depends on the number of alternative values contained inside each uncertain tuple. When these values are close to each other the sort-based approach is almost as efficient as a traditional sort-join. In addition the sort-based approach improves significantly when we require only the first few tuples from the result, as it usually happens in modern database system GUIs where additional tuples are fetched into memory only if explicitly required by the user. However, in this case using an index is the most efficient approach. On all our experiments the algorithm chosen by the underlying relational system, which is not aware of the uncertainty, is outperformed by at least one of the uncertainty-aware approaches, and using the most appropriate algorithm we have been able to efficiently join both synthetic and real data and tables containing millions of uncertain tuples.

\bibliographystyle{plain}
\bibliography{BIB-Bibliography,magnani}

\end{document}